\pdfoutput=1
\documentclass[]{aa}
\usepackage{natbib}

\newcommand{\apropto}{\;
  \raise0.3ex\hbox{$\propto$\kern-0.75em\raise-1.1ex\hbox{$\sim$
  }}\;\hskip-2pt }
\usepackage{graphicx}

\newcommand{\lta}{\;
  \raise0.3ex\hbox{$<$\kern-0.75em\raise-1.1ex\hbox{$\sim$
  }}\;\hskip-2pt }
\newcommand{\gta}{\;
  \raise0.3ex\hbox{$>$\kern-0.75em\raise-1.1ex\hbox{$\sim$
  }}\;\hskip-2pt }

\begin{document}
\title{Reversals of the solar magnetic dipole in the light of observational data and simple dynamo models
}

   \author{D.~Moss\inst{1} \and V.~V.~Pipin\inst{2} \and D.~Sokoloff\inst{3} \and J.~T.~Hoeksema\inst{4}}

   \offprints{D.Moss}

   \institute{School of Mathematics, University of Manchester, Oxford Road,
Manchester, M13 9PL, UK \and
Institute of Solar-Terrestrial Physics RAS, Irkutsk 664033, Russia
\and
Department of Physics, Moscow University, 119992 Moscow, Russia
\and
  W.W.Hansen Experimental Physics Laboratory, Stanford University, Stanford, CA 94305, USA}

   \date{Received ..... ; accepted .....}

\abstract{Observations show that the photospheric solar magnetic dipole usually does not vanish during the reversal of the solar magnetic field,
which occurs in each solar cycle. In contrast, mean-field solar dynamo models
predict that the dipole field does become zero.
In a recent paper Moss et al. (2013) suggested that this contradiction 
can be explained as a large-scale 
manifestation of small-scale magnetic fluctuations of the surface poloidal field.}
{Our aim is to confront this interpretation with the available observational data.}
{Here we compare this interpretation with WSO (Wilcox Solar
Observatory) photospheric magnetic field data in order to determine the amplitude of magnetic fluctuations 
required to explain the phenomenon and to compare the results with predictions 
from a simple dynamo model which takes these
fluctuations into account.}{We demonstrate that the WSO  data concerning the magnetic dipole reversals are very 
similar to the predictions of our very simple solar dynamo model,
 which includes both  mean magnetic field and 
fluctuations. The ratio between the rms value of the magnetic fluctuations 
and the mean field is estimated to be about 2,  
in reasonable agreement with estimates from sunspot data. 
The reversal epoch, during which the
fluctuating contribution to the dipole is larger than that from the mean field, 
 is about 4 months. The memory time of the 
fluctuations is about 2 months.  Observations demonstrate that the rms of the
 magnetic fluctuations is strongly modulated by the phase of the
solar cycle. This gives additional support to the concept that the
solar magnetic field is generated by a single dynamo 
mechanism rather than also by independent small-scale dynamo action. 
A suggestion of a weak nonaxsymmetric magnetic 
field of a fluctuating nature arises from the analysis, with a lifetime of
about 1 year.
} 
{The behaviour of the magnetic dipole during the reversal epoch gives valuable information about details of 
solar dynamo action.}

\keywords{magnetic fields -- magnetic fields -- Sun: magnetic fields -- Sun:
activity}

\titlerunning{Reversals of the solar magnetic dipole}
\authorrunning{Moss et al.}

\maketitle

\section{Introduction}

\label{intro}

\begin{figure}
\includegraphics[width=0.99\columnwidth]{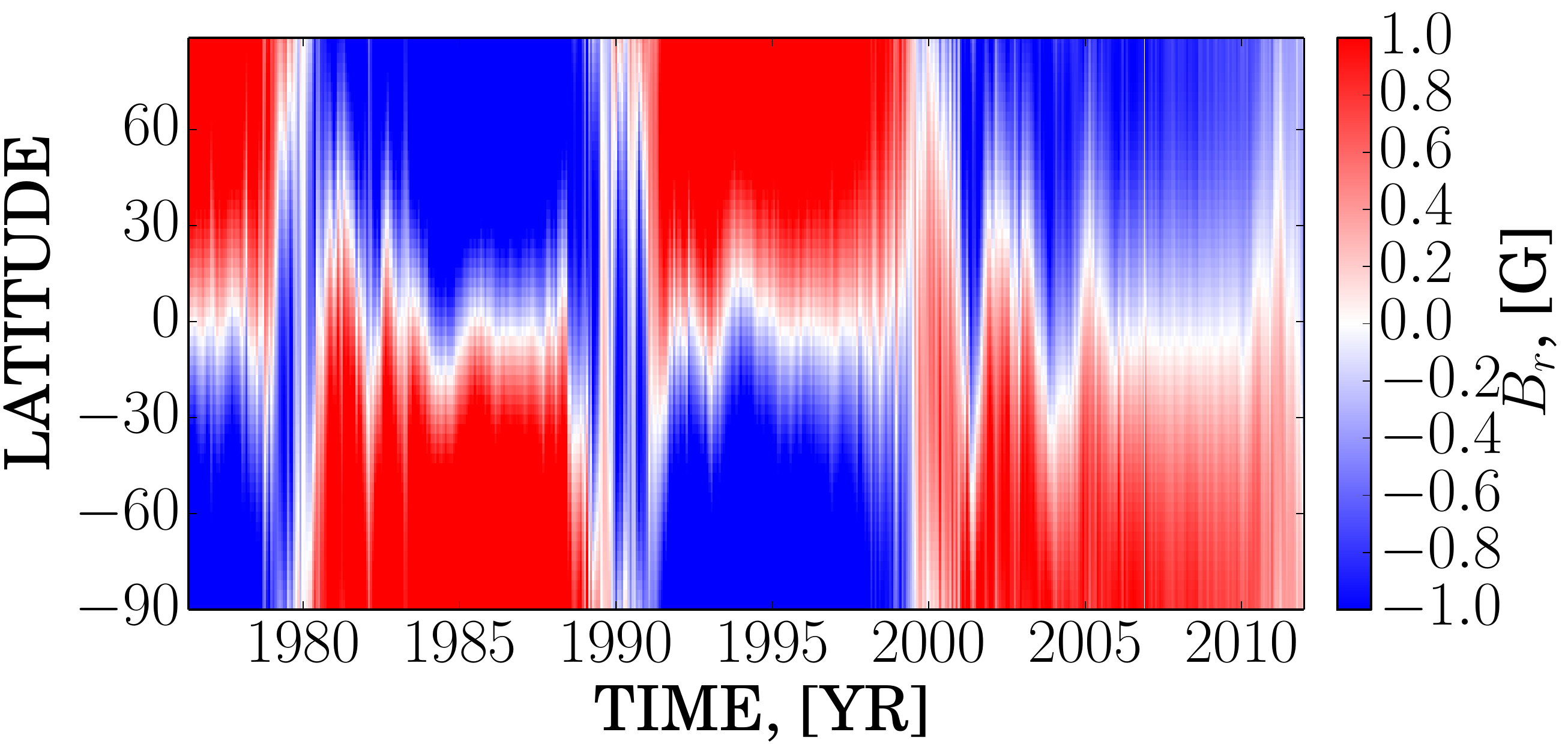} 
\caption{Evolution of the dipole component of the radial magnetic
  field as obtained from measurements at the  Wilcox Solar
  Observatory (WSO) from 1976 to 2013.}
\label{inv1} 
\end{figure}

Our research presented in this paper starts from the 
following quite simple statements. The photospheric
solar magnetic dipole reverses its orientation every 11 years, in
the course of the solar activity cycle.  It is  believed that the solar
cycle is driven by a dynamo mechanism operating deep  in the convection zone. 
The evolution of the solar dipole is synchronized with sunspot activity.
Details of this synchronization may depend on the underlying dynamo
mechanism \citet{PK13}. Fig.~\ref{inv1} shows
reversals of the dipolar component of the large-scale
magnetic field of the Sun.

The Sun is basically axisymmetric so a natural expectation is that the solar magnetic
field is also approximately axisymmetric. Correspondingly, a standard
mean-field description of the solar dynamo assumes axial symmetry and
that the solar dipole vanishes at the instant of
reversal. 
In contrast, observational data (Livshits \& Obridko 2006; DeRosa
et al. 2012) show that the solar magnetic dipole rotates from pole
to pole during the course of a reversal, even becoming orthogonal
to the rotation axis at some instant,  corresponding obviously to the vanishing of
the polar dipole. At such a time the  total
magnetic field is strongly nonaxisymmetric.

Moss, Kitchatinov \& Sokoloff (2013) suggested that this apparent
contradiction is explicable as a manifestation of contributions from
{ fluctuations of the surface magnetic field (specifically
its poloidal component)} near the moment of reversal. The aim of this
paper is to develop and present a quantitative model that allows direct
comparison of the interpretation of Moss et al. (2013) with observations.
Using a simple simulation of this process, we produce models in
which the trajectory of the dipole axis on the surface mimics well
the observed behaviour. Other properties of the model also resemble
those observed.

In this paper, we concentrate on the fact (known for some time from solar activity studies, but little exploited) 
that the large-scale magnetic field becomes substantially
nonaxisymmetric during a reversal. Quantifying this nonaxisymmetry
in terms of { fluctuations of the surface solar magnetic field}, we suggest
that something can be learned about the relative contributions of mean-field 
and small-scale dynamos to the solar magnetic field (Sect.~\ref{disc}). We note
that the relationship between dipolar and quadrupolar
modes in solar dynamo action was addressed by DeRosa et al. (2012)
and we do not discuss it here. The manifestation of a substantial nonaxisymmetric 
component of the solar dipolar magnetic field is interpreted in the framework 
of mean-field dynamo theory.


One further point is that the apparent discrepancy between dynamo theory
and the observations 
applies explicitly (in the form discussed above) only
to mean-field dynamos.
Of course, the solar magnetic field obtained from direct numerical
simulations (DNS) is not strictly axisymmetric -- even for hydrodynamics which
is axisymmetric on average. In principle, one might prefer to use
DNS and thus avoid any mean-field interpretation
with its associated uncertainties,
rather than involving magnetic fluctuations in the mean-field description.
Of course, at the moment DNS models 
are expensive in terms of computing resources required.
A hint that the sort of behaviour studied here 
may not be incompatible with DNS is given by Brown et al. (2011),
{ where one model appears to show a latitudinal migration of the dipolar
component} .
We believe that the mean-field formulation of dynamo theory
is a fruitful way to obtain a qualitative understanding of the solar cycle,
and then compare it with observations and formulation. 
Thus the formulation of an adequate description
of the mean-field behaviour of the solar cycle is a desirable undertaking.

One outcome of the investigation presented here is  the recognition
that the observational analysis isolates 
some features of the solar dynamo that appear instructive and informative
in various other contexts related to solar dynamos. We discuss 
such byproducts at the end of the paper (Sect.~\ref{disc}).

\section{Methodology}
\label{method}

Our tactic in comparing theory and observations is as follows.
First we must understand the capability of a
mean-field dynamo model in explaining the phenomenon under discussion.
A reasonable approach is to take the simplest dynamo model, compare
it with observations and then add more and more realistic details until
the desired agreement is achieved. We start with the illustrative
Parker (1955) dynamo model and find that it is adequate to provide
the mean-field component of the theoretical modelling. This makes using
a detailed mean-field model unnecessary at this stage. Of course, 
we recognize the purely illustrative
nature of this toy model (for example, the incorrect phase relation between 
toroidal and poloidal magnetic fields); however such issues are 
not relevant for the point being
discussed. Note that the primitive nature of this mean-field
model allows us { to introduce a parameter which quantifies the 
link between}
toroidal and poloidal magnetic field and so
avoid any specific connection with the $\alpha$-effect introduced
by Steenbeck, Krause and R\"adler (see Krause and R\"adler  1980). Thus
the results should also be applicable to flux-transport
models (e.g. Choudhuri  2008).

Standard mean-field dynamo models do not contain explicit magnetic fluctuations
and we have to include them somehow.
This can be done in various forms and again we do
it in the simplest way, i.e. by adding magnetic fluctuations "by hand"
using a random number generator,  as described in Sect.~\ref{dyn}. 
This is found  to be sufficient to reproduce the observational phenomenology.
The reversal phenomenon is quite robust and does not
depend on details of the model. We see below that
the magnetic fluctuation properties from our analysis
are meaningful for understanding the surface manifestations 
of solar dynamo action.
Thus there is no motivation for considering more realistic
dynamo models.
Of course the true solar dynamo is much more
complicated than the toy model we use  to mimic the observations.

Various approaches have been discussed for
  determining the position of the solar magnetic pole, see, for example,
\cite{Kukl88} and \cite{pat2013}.
A delicate feature of our analysis is the fact that the observational
data exploited are observations of the magnetic field at the solar surface.
Such a field is mainly contributed by the magnetic fields that originate in solar active
regions, which in turn are mainly determined by the toroidal magnetic field
within the Sun. On the other hand, we are interested in the solar dipole
magnetic moment and the toroidal field gives no contribution to this quantity.
Of course, the fact that only a tiny part of the surface magnetic
field determines the quantity being discussed requires some care
in using the data. We note however that the situation is
recognized by solar observers 
who have verified the reliability of the result with great care.

\begin{figure}
\includegraphics[width=0.95\columnwidth]{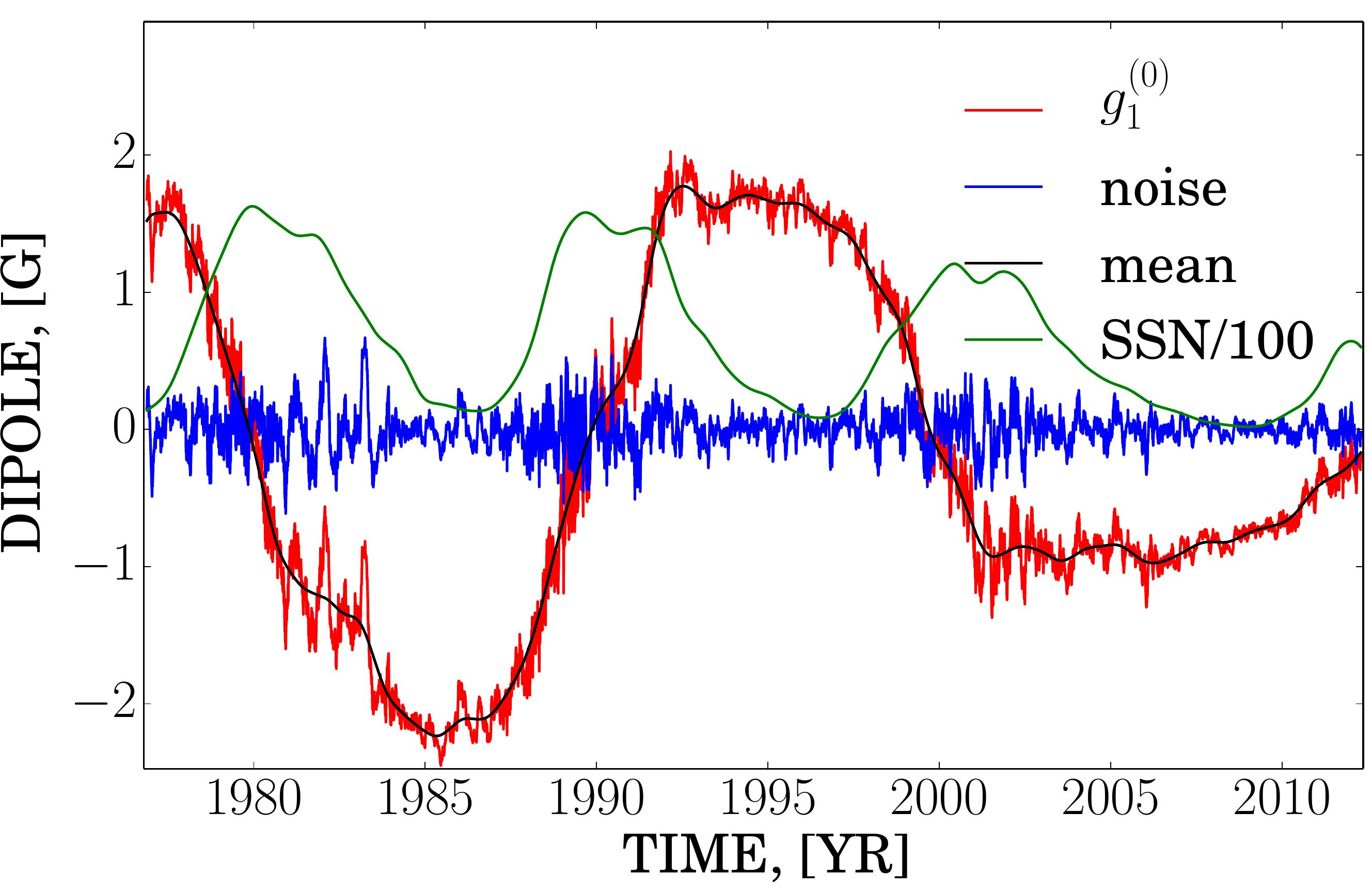} 
\caption{Temporal evolution of the axisymmetric component of the solar magnetic
dipole (red), its Gaussian filtering with a 1-year window (black) and
the residual noise (blue), compared with sunspot number.}
\label{dip1} 
\end{figure}

\section{Observational data}
\label{obs}
{ In the paper we analyze the components of the global magnetic field 
deduced from the daily  magnetograph observations that have been made
at the Wilcox Solar Observatory (WSO) at Stanford since May of
1976. The solar magnetograph measures the line-of-sight component of
the photospheric magnetic field  with three arc min resolution (see
details in \citealp{Sch77} and \citealp{du77}). 
This homogeneous data set provides information about the evolution of the photospheric
magnetic field through the past three Solar Cycles, 21, 22 and 23. The
daily magnetograms were interpolated on to the Carrington grid and 
assembled into synoptic charts which were further used to determine
the coefficients of the spherical harmonics of the potential magnetic
field outside of the Sun. The procedure is described in detail by Hoeksema \& Scherrer (1986). 
Below, we discuss some particular points which are essential for the
current study.}

\begin{figure}
a)\includegraphics[width=0.85\columnwidth]{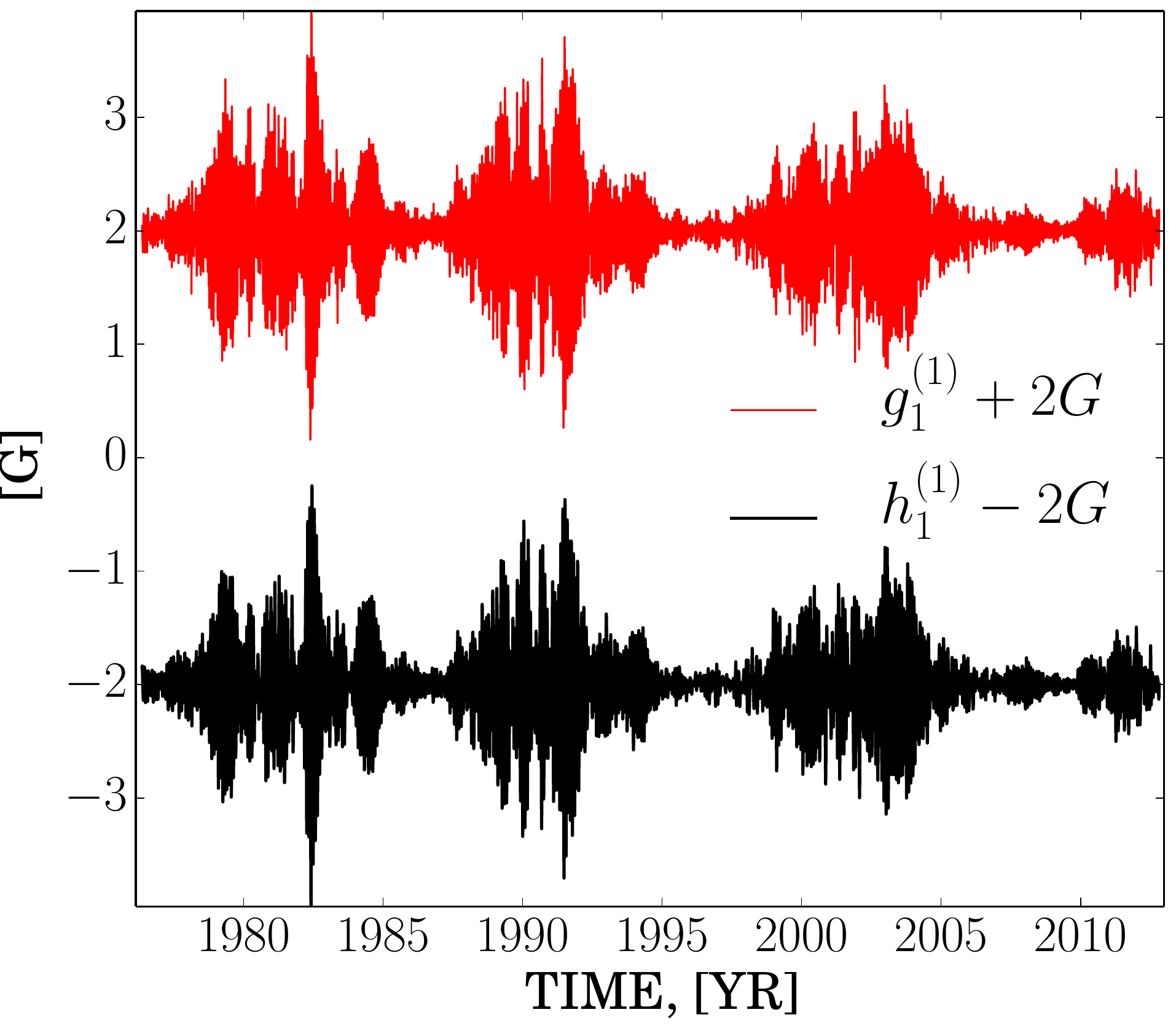}\\ 
b)\includegraphics[width=0.95\columnwidth]{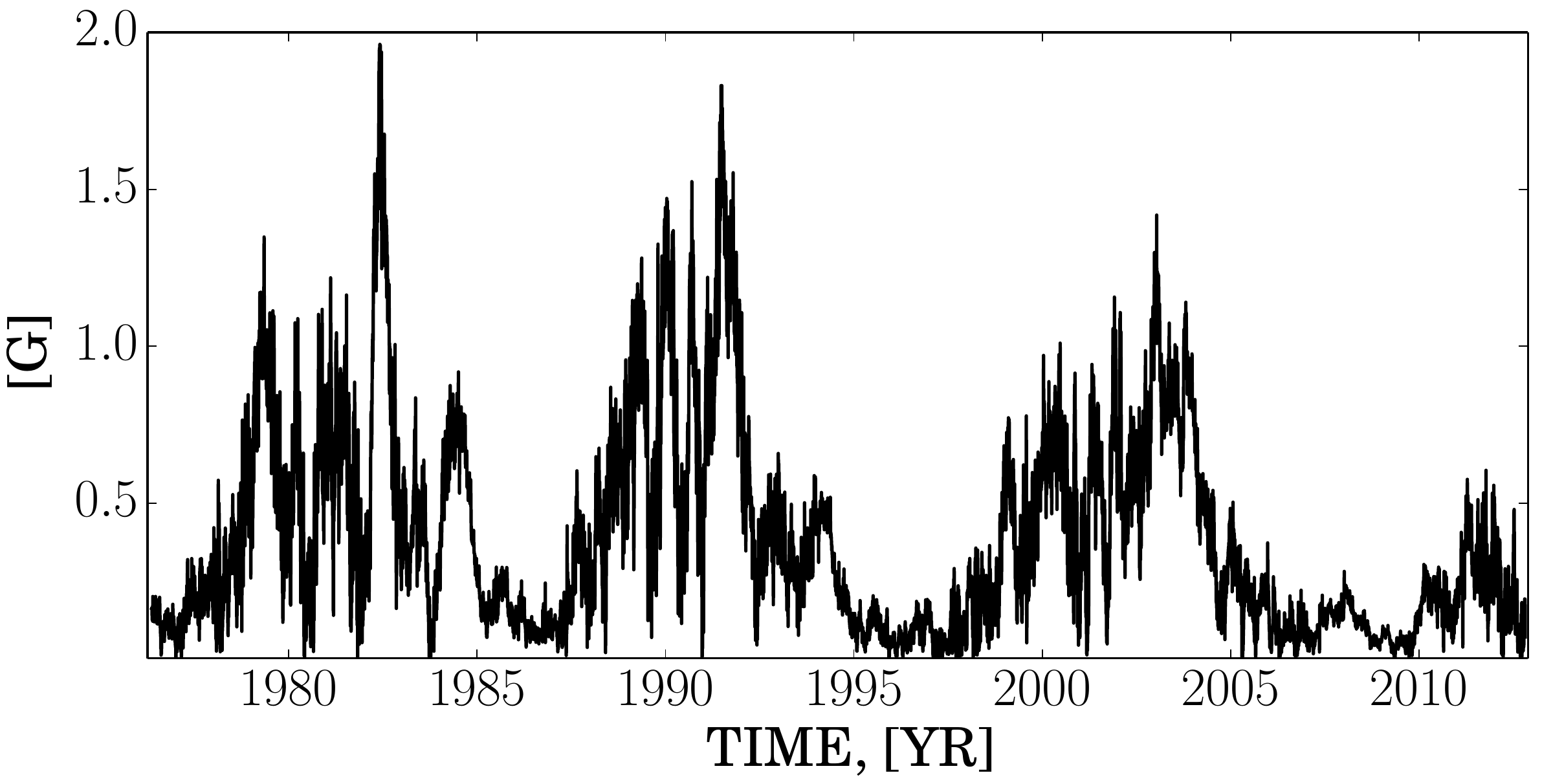}\\
\caption { a) The components of the solar equatorial dipole; b) the amplitude of the equatorial dipole.}
\label{orig} 
\end{figure}

We consider the spherical harmonic decomposition of a scalar potential
\begin{eqnarray}
\psi=R\sum_{n=1}^{\infty}\sum_{m=0}^{n}\left(\frac{R}{r}\right)^{n+1}[g_{n}^{(m)}\cos\left(m\phi\right)+\nonumber \\
+h_{n}^{(m)}\sin\left(m\phi\right)]P_{n}^{m}\left(\theta\right).
\end{eqnarray}

The potential magnetic field above the solar surface can be represented as a gradient
of a scalar potential, ${\mathbf B}=-{\mathbf \nabla}\psi$. 
The data reduction uses the calculations of the radial harmonic coefficients
of the photospheric magnetic field described by \cite{zhhx93},
see also .
Hoeksema \& Scherrer (1986). { Note, that  the fitting procedure is
  made using the radial magnetic field as a boundary condition (see, \citealp{zhhx93})}.
 Coefficients of the spherical harmonics
are computed for each 10 degrees of Carrington rotation. In our analysis
we use the components of the axial and equatorial dipoles provided
by the first three coefficients of the potential field extrapolation
of the photospheric magnetic field, namely, $g_{1}^{(0)}$, $g_{1}^{(1)}$
and $h_{1}^{(1)}$. 
These three coefficients give the dipole part
of the radial component of magnetic field:

\begin{eqnarray}
B_{r}=2\left(\frac{R}{r}\right)^{3}[g_{1}^{(0)}\cos\theta-(g_{1}^{(1)}\cos\phi+\\
+h_{1}^{(1)}\sin\phi)\sin\theta].\label{Br}
\end{eqnarray}
{ From these three coefficients we define the inclination of the effective
dipole and its azimuth by

\begin{eqnarray}
g^{(h)} & = & \sqrt{ {g_{1}^{(1)}}^2+{h_{1}^{(1)}}^2}, \label{amp}\\
\Theta & = & \arctan\left(\frac{g_{1}^{(0)}}{g^{(h)}}  \right),\label{tilt}
\end{eqnarray}
where $\Theta$ is the inclination of the dipole, and $g^{(h)}$ is amplitude of the equatorial
dipole. 
The components $g_{1}^{(1)}$ and $h_{1}^{(1)}$ determine the Carrington longitude  of the
equatorial dipole as $\arctan\left(h_{1}^{(1)}/g_{1}^{(1)}\right)$,
its azimuth $\Lambda$ is calculated as a sum of the Carrington longitude
and the  phase offset,  $\Phi_{{\rm Car}}$.}
{ Figures \ref{dip1} and \ref{orig}a,b show the evolution of the components of the dipolar
field of the Sun.}

\section{A dynamo model}

\label{dyn}

Our idea is to start with a simple axisymmetric
dynamo model and to add small-scale nonaxisymmetric injections of
poloidal magnetic field, distributed randomly in area
over the whole surface  of
the sphere. For the underlying (axisymmetric) dynamo model we use
an extension of the 1D Parker model, as described by Moss, Sokoloff,
Kuzanyan \& Petrov (2004){, see also Moss, Saar \& Sokoloff (2008)}. 
The governing equations for the toroidal
field $B(\theta)$ and potential $A(\theta)$ for the poloidal field are

\begin{equation}
\frac{\partial B}{\partial t}=Dg\sin\theta\frac{\partial A}{\partial\theta}+\frac{\partial^{2}B}{\partial\theta^{2}}-\mu^{2}B,
\label{eqB}
\end{equation}
\begin{equation}
\frac{\partial A}{\partial t}=\alpha B+\frac{\partial^{2}A}{\partial\theta^{2}}-\mu^{2}A.
\label{eqA}
\end{equation}

Here $\theta$ is the polar angle,
the factor $\mu$ is introduced to represent radial diffusion, and
we take $\mu=3$ as appropriate for a dynamo region of about $30\%$
in radius. The factor $g(\theta)$ is
introduced to allow a representation
of latitudinal variations in the rotation law -- we take $g=1$ -- and
take the dynamo number $D=-10^{3}$. Of course it would be possible
to use a more sophisticated, two-dimensional, dynamo model. However
we are not here studying details of the overall field behaviour as
a function of radius, but rather we only need a mechanism to
produce approximately sinusoidal variations of radial field at the
surface.
\begin{figure}
a)\includegraphics[width=0.94\columnwidth,height=0.16\paperheight]{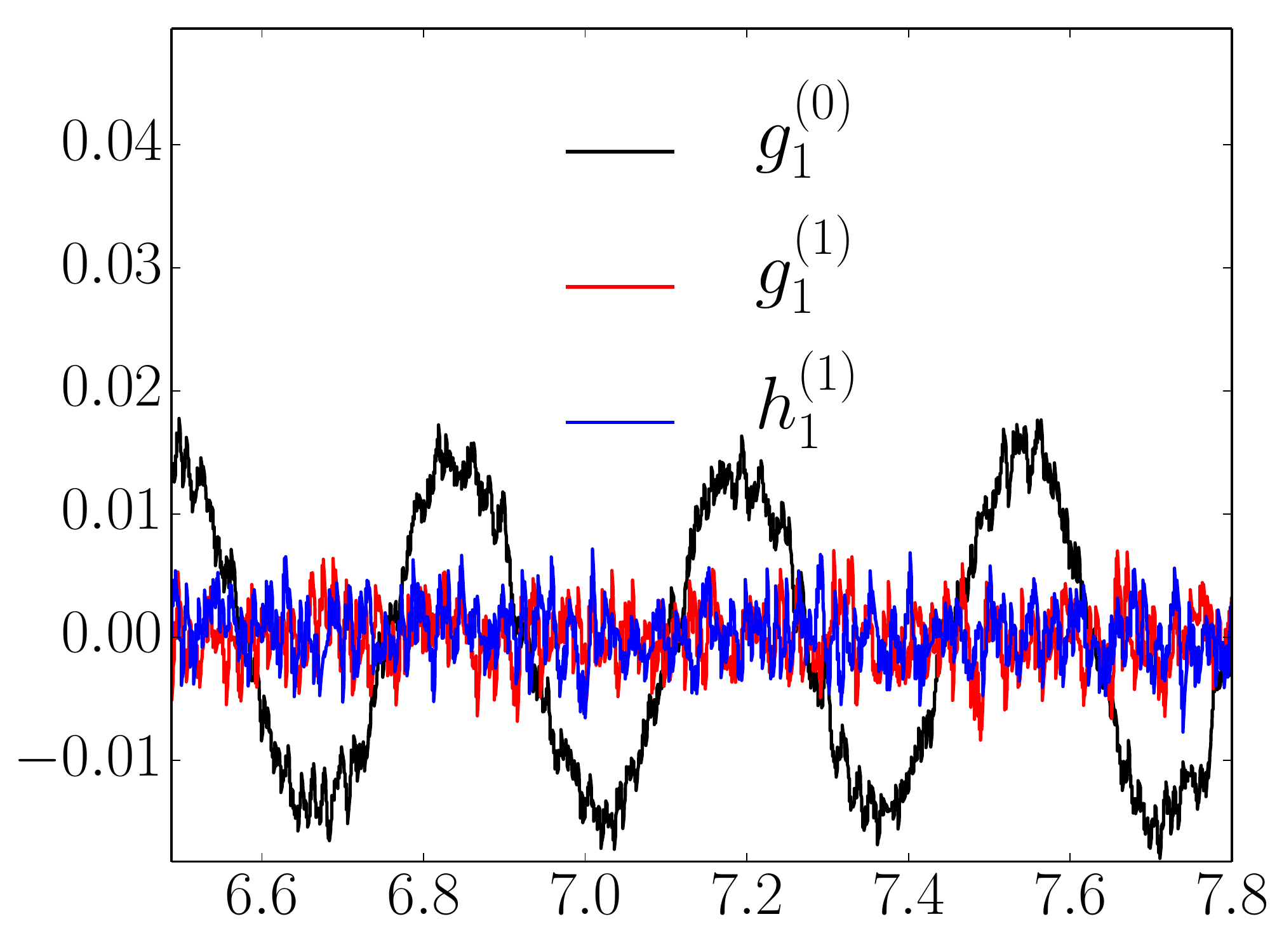} 
b)\includegraphics[width=0.95\columnwidth,height=0.16\paperheight]{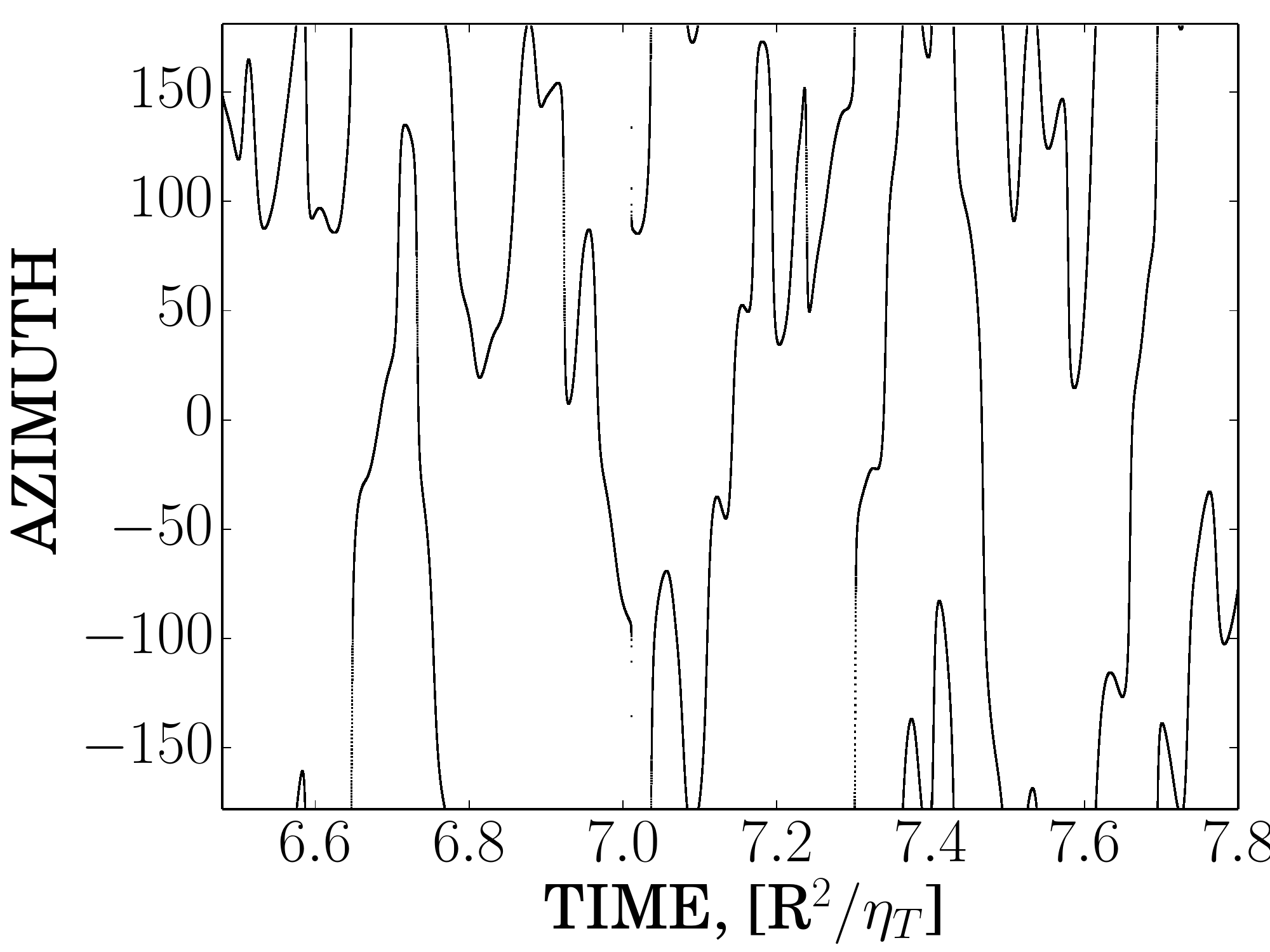} 
\caption{a) Evolution of the magnetic dipole  components from the model
  data (with time independent  poloidal field injection amplitude
of arbitrary sign). b)  Evolution of azimuth (see comments in the text).
{ If the half-period (0.176) of the oscillations in panel a) is identified
with the 11 yr sunspot cycle, then the time unit is about 62.5 yr.}}
\label{mod} 
\end{figure}

Once our model has settled to a regular oscillatory behaviour 
{ we map the axisymmetric field from the mean field dynamo on to
the surface of a sphere. Then} we add,
at fixed time intervals $dt_{{\rm inj}}$,  co-rotating patches of radial field,
strength $B_{{\rm inj}}$ { of arbitrary sign}, 
in circular regions at randomly located
positions on the surface of the sphere. In the models described, these
regions are of radius $20-30^{\circ}$. They are intended to represent
injections of field from solar active regions - we are only interested
in the resulting contributions to the global poloidal field. 
{The resulting global poloidal field is thus nonaxisymmetric,
being the sum of the symmetric dynamo generated field and the injections which 
are random in longitude (and in latitude).
In our modelling the injected field is just added formally to the poloidal 
field output from Eqs.~(\ref{eqB}), (\ref{eqA}) -- the injected field is not 
input back into the dynamo equations. We can justify this procedure by noting
that the dynamo is the result of processes occurring deep within the Sun, and
can hardly be affected by the superficial phenomena associated with
surface activity ("the tail does not wag the dog").}

The injected field has strength of order the maximum dynamo poloidal field
and decays with time constant $t_{{\rm dec}}$. The code allows multiple
spots to coexist, with each generated successively after the fixed
interval $dt_{\rm inj}$.  If appropriate, we can modulate the injection 
amplitude with the phase of solar cycle.
The evolution of the dipole components { for a typical realization of
 the model} is illustrated in
Fig. \ref{mod}a.  
The result is not very sensitive to the input parameters. 
The azimuth of the dipole is random,
because of the nature of the field injection mechanism.
{ It is determined by the position of the instantaneous dipole, which
in turn is given by the most recent surface field fluctuations.
The evolution of the toroidal field  follows the evolution of the axial dipole
with phase shift of about $\pi/2$,
and with amplitude of the toroidal field larger by a factor of about 40 compared to
the poloidal. (This ratio could easily be adjusted by changing the parameters
in Eqs.~(\ref{eqB}), (\ref{eqA}), without changing our general conclusions.)}
Fig. \ref{mod}b illustrates the evolution of azimuth, smoothed using a
Gaussian filter with a window which corresponds to one year in the model.
{Any apparent correlation in the evolutionary paths in Fig. 6 appears to be coincidental.}

\section{Results}

\subsection{Basic analysis of the data}
\label{basic}

\begin{figure}
\includegraphics[width=0.95\columnwidth]{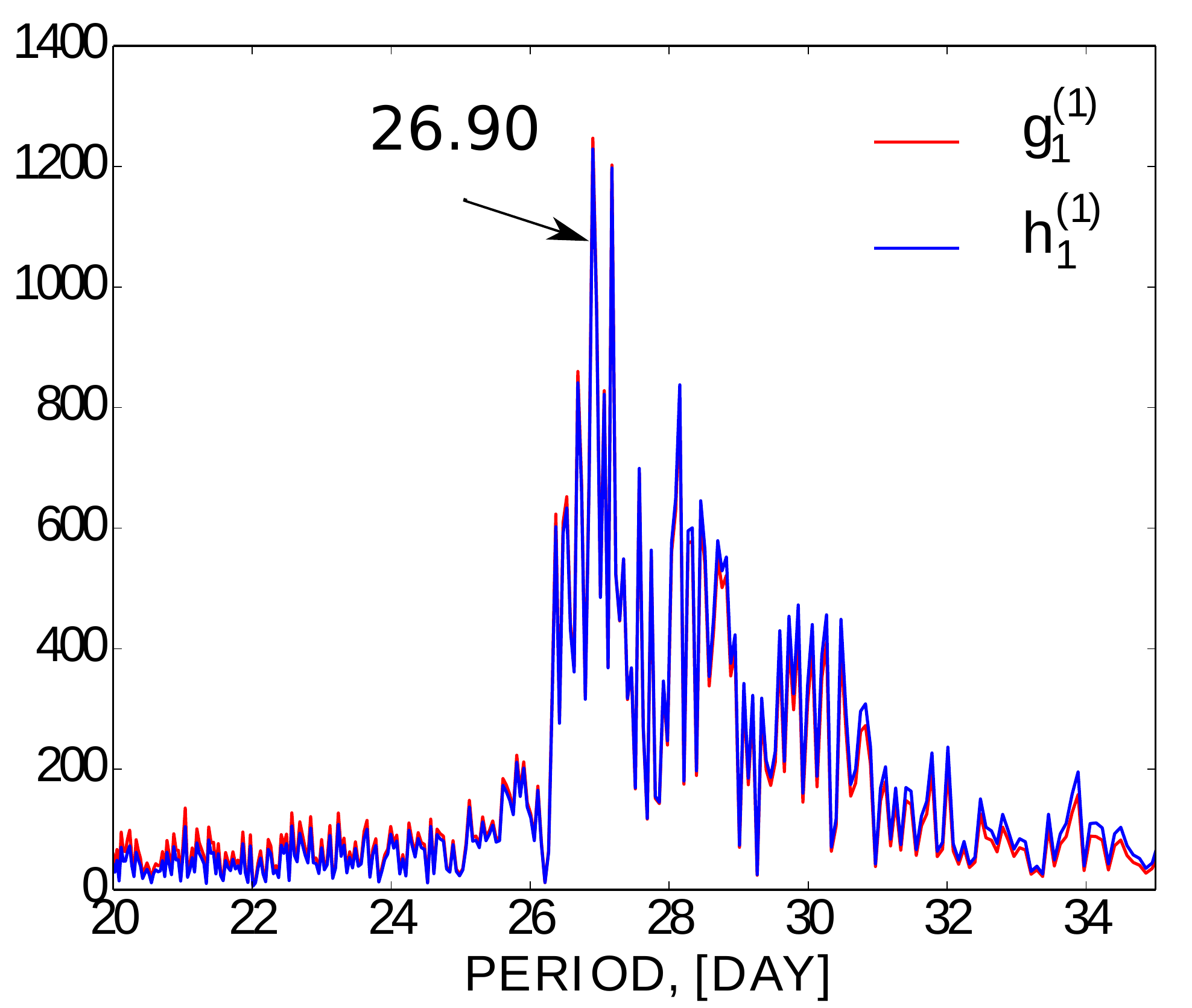} \caption{Spectrum of the equatorial dipole components.}
\label{spek} 
\end{figure}

For the purpose of analysis we have to separate the mean and fluctuating
parts of the field and then do the same for the inclination and azimuth of
the dipole. For the axisymmetric part of the dipole, which is represented
by $g_{1}^{(0)}$, this can be done by suitable filtering
of the data in the time series. We use a Gaussian filter with a one
year window, which is often used in analysis of the sunspot data
set \cite{hath09}. Fig.~\ref{dip1} shows the evolution of 
$g_{1}^{(0)}$, its mean and noise components, and the smoothed sunspot
number as given by the \cite{sidc} data base.  
{ Note that the noise is {\em not} the fluctuating part of the field (as defined in Sect.~\ref{dyn}),
but the numerical residual after data analysis.}

\begin{figure}
a)\includegraphics[width=0.85\columnwidth]{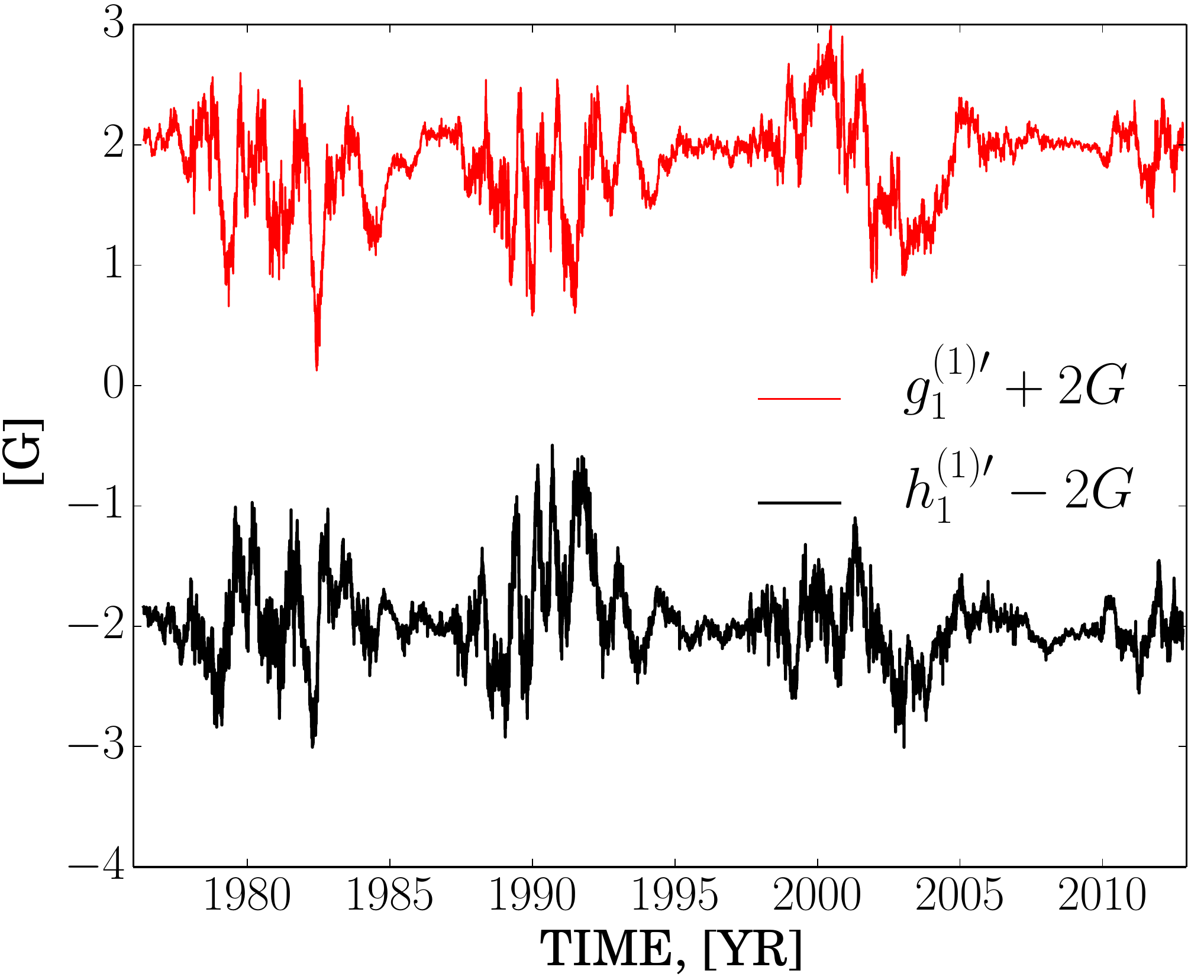}
b)\includegraphics[width=0.95\columnwidth]{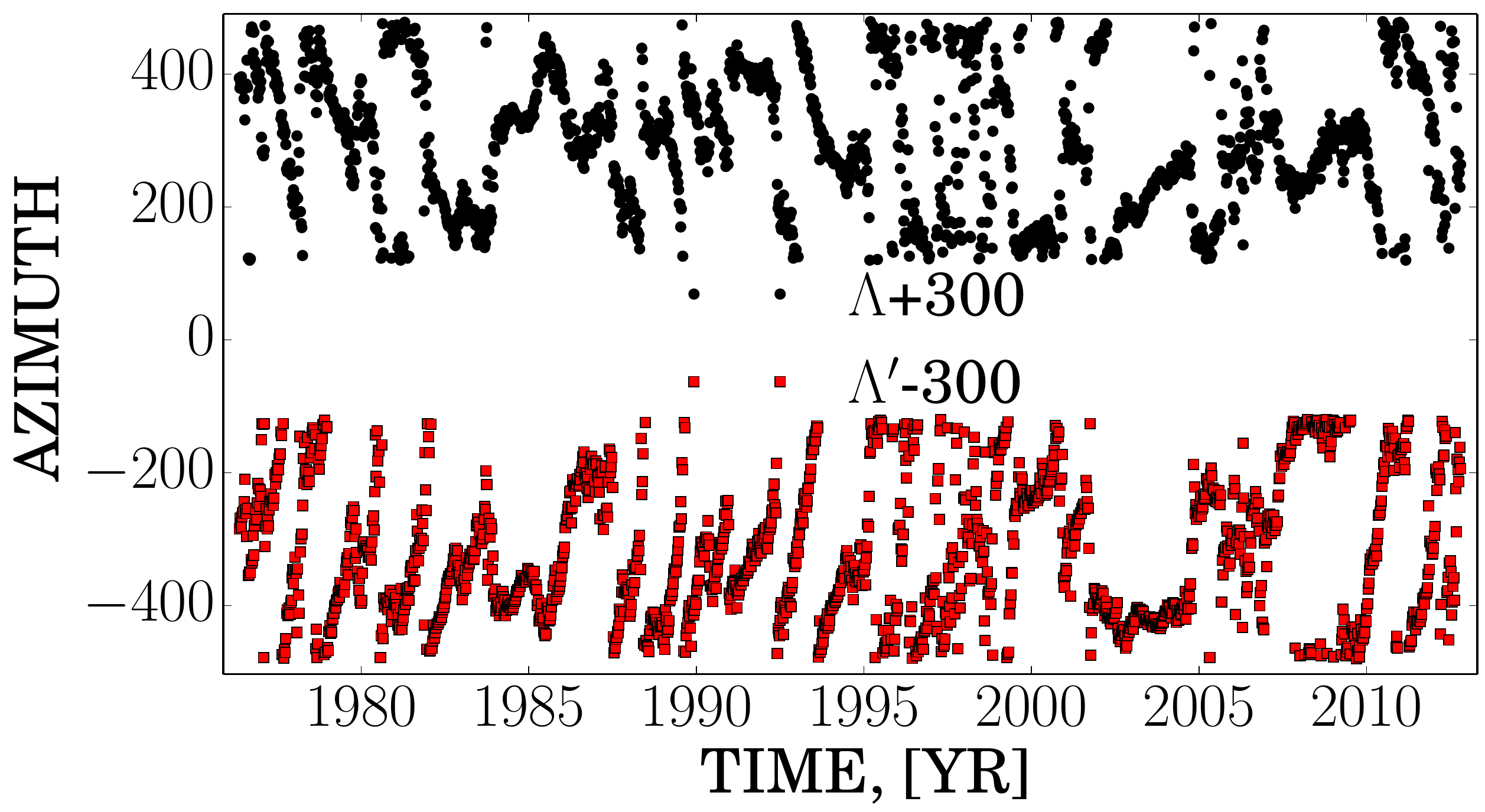}
\caption{a) The  transformed components of the equatorial dipole,
rotation period $P=26.90$ days; b)
 the upper curve shows the azimuth of the equatoial dipole
in the frame rotating with the Carrington period P=27.253 days,
the lower curve shows the azimuth for the mode P=26.90 days.}
\label{transf} 
\end{figure}

{ Now we move to the analysis of the behaviour of the equatorial dipole, and our first aim is to determine the 
rotation rate of this dipole. For this purpose we determine Fourier spectra  
of $g_{1}^{(1)}$ and $h_{1}^{(1)}$  using the standard 
fast Fourier transform (FFT) routines provided by Scientific Python (www.scipy.org). 
Fig.~\ref{spek} shows the result. Very similar spectra were found earlier
by  Svalgaard \& Wilcox (1975) and \cite{kotov83} from analysis of the rotation of the interplanetary magnetic
field. We see that the spectra have peaks in a range of synodic rotation periods between 26 
and 30 days, including the Carrington period of 27.2753 days. We recall that the synodic rotation period  refers to
motion as observed from Earth. Our interpretation of the plot is that the azimuthal dipole field is, of course, not 
just white noise and that its memory is sufficiently long to feel the 
dipole rotation and to identify at least the main 
peak in the spectra, which corresponds to the rotation period $P=26.90$ day.
The scatter of the periods in 
Fig.~\ref{spek} could be a consequence of the differential rotation. In this paper
we restrict our analysis to a particular mode with period $P=26.90$ day.

We perform a filtering of the azimuthal dipole data in the frame rotating with the period $P=26.90$ day using the following transformation}

\begin{figure}
\includegraphics[width=0.95\columnwidth]{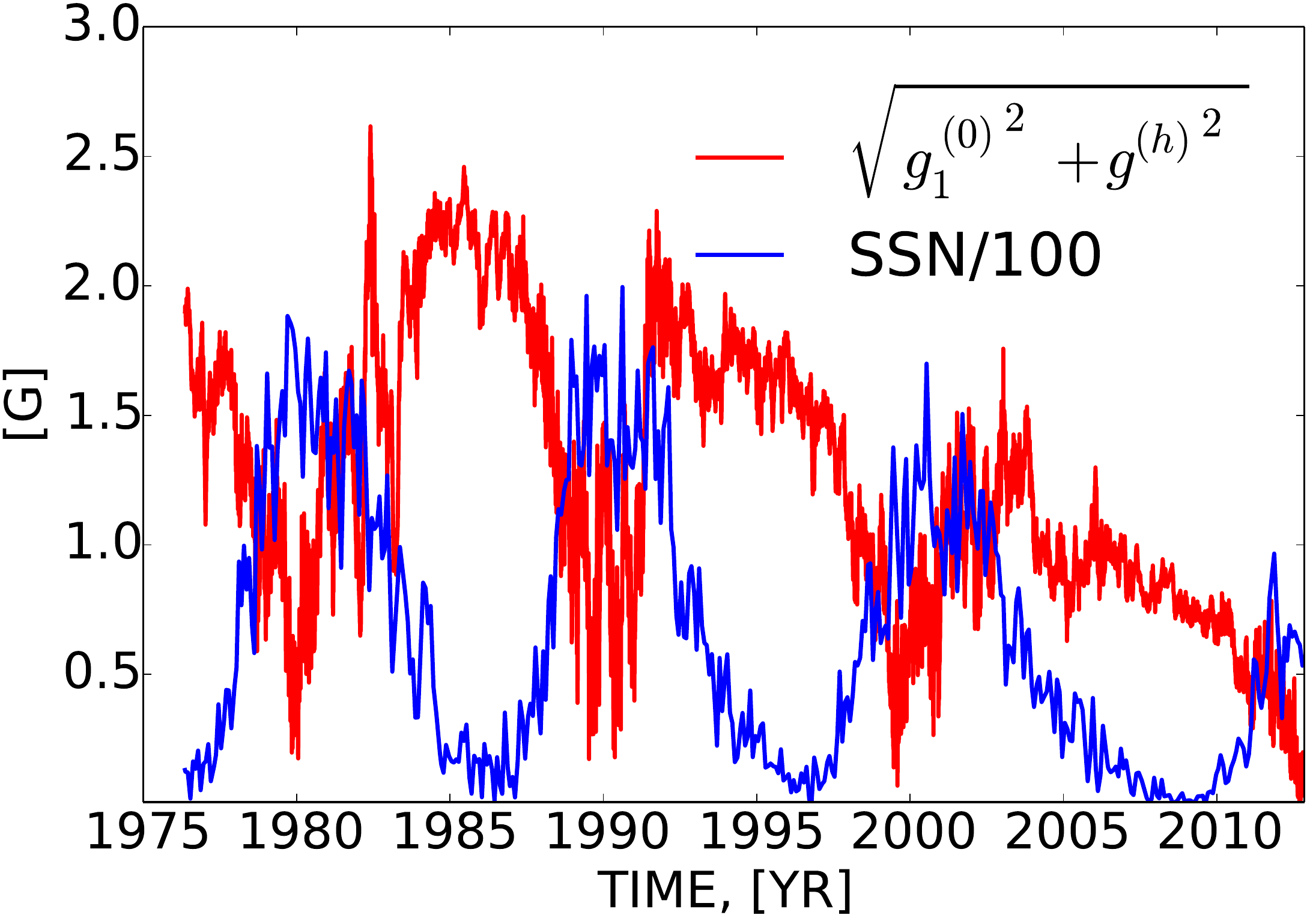} 
\caption{Evolution of the large-scale dipolar magnetic field components plotted with the sunspot number data. 
\label{datam}}
\end{figure}

\begin{eqnarray}
g_{1}^{(1)\prime} & = & g_{1}^{(1)}\cos\Omega t+h_{1}^{(1)}\sin\Omega t\label{transfe}\\
h_{1}^{(1)\prime} & = & h_{1}^{(1)}\cos\Omega t-g_{1}^{(1)}\sin\Omega t.\nonumber
\end{eqnarray}
{ Fig.~\ref{transf}a shows  evolution of the components obtained in result of this transformation} for the
case $\Omega=2\pi/P$, $P=26.90$ day. { Of course, similar }results can be obtained
for other periods from the given range.
The amplitude of the equatorial dipole remains the same under
transformation.

{ Fig.~\ref{transf}(b) shows how different the azimuth can be if we
follow the original equatorial dipole components (Fig.~\ref{orig})
and the components of the mode  $P=26.90$ day. For example,
we observe the quasi-regular large-scale fluctuation of this mode
during the descending phase of cycle 23, when the mode was permanently
visible around longitude $260^\circ$  for a period of about 3 years
{\bf starting around the  year 2002.} 
{\bf The existence of such events is also suggested by results of thorough
period analysis (see, e.g., Berdyugina et al 2006)}
We have to stress that the 
our particular procedure of analysis is employed because of our restriction to
consideration of the first three modes of the decomposition given by
the  definitions of azimuth from Eq.~(1). 
The azimuth can be uncertain if the signal has a complicated spectrum
like that shown by Figure \ref{spek}. 
In the literature, the reader can find some
alternative possibilities for determination of the azimuth of the solar
dipole (e.g. \cite{Kukl88, pat2013}). However, our method allows direct
comparison of  results with predictions of the dynamo models.}

The nonaxisymmetric
modes have the largest variations in time. Using the modes in
the rotating coordinate frame we will define the inclination and the azimuth
of dipole for the same reference frame. For the remainder of our
illustrations we use the mode corresponding to the period 
$P=26.90$ day. This value corresponds  to the position of the highest peak
in  Fig.~\ref{spek}. Clearly, in our definition the
azimuth of the global dipole has an arbitrary zero point.
On the other hand, it is well-known that the given characteristics of
the large-scale magnetic field of the Sun trace the sectorial structure
of the interplanetary magnetic field in the heliosphere
\cite{WSO2}. A problem in determination of the global dipole azimuth is that the two-sector structure, which is observed during
much of the 
cycle, changes to a four sector structure
during the maxima of the sunspot  cycle. The sectorial structure
reflects the total contribution of all the non-axisymmetric modes. It
can be used for the azimuth determination only during the phases when the 
two-sector structure dominates. 

{ We show the long-term evolution of the magnitude of the total dipole 
strength of the solar field in Fig.~\ref{datam},
together with the sunspot number data. 
The Sun currently appears to be presenting rather atypical behaviour
in the context of the last century or so. The ongoing
decline in activity emphasizes that we are not dealing with a strictly periodic
system and that our modelling, by assuming an underlying regular 
periodicity, can
only hope to give a partial representation.}

\begin{figure}
\includegraphics[width=0.9\columnwidth]{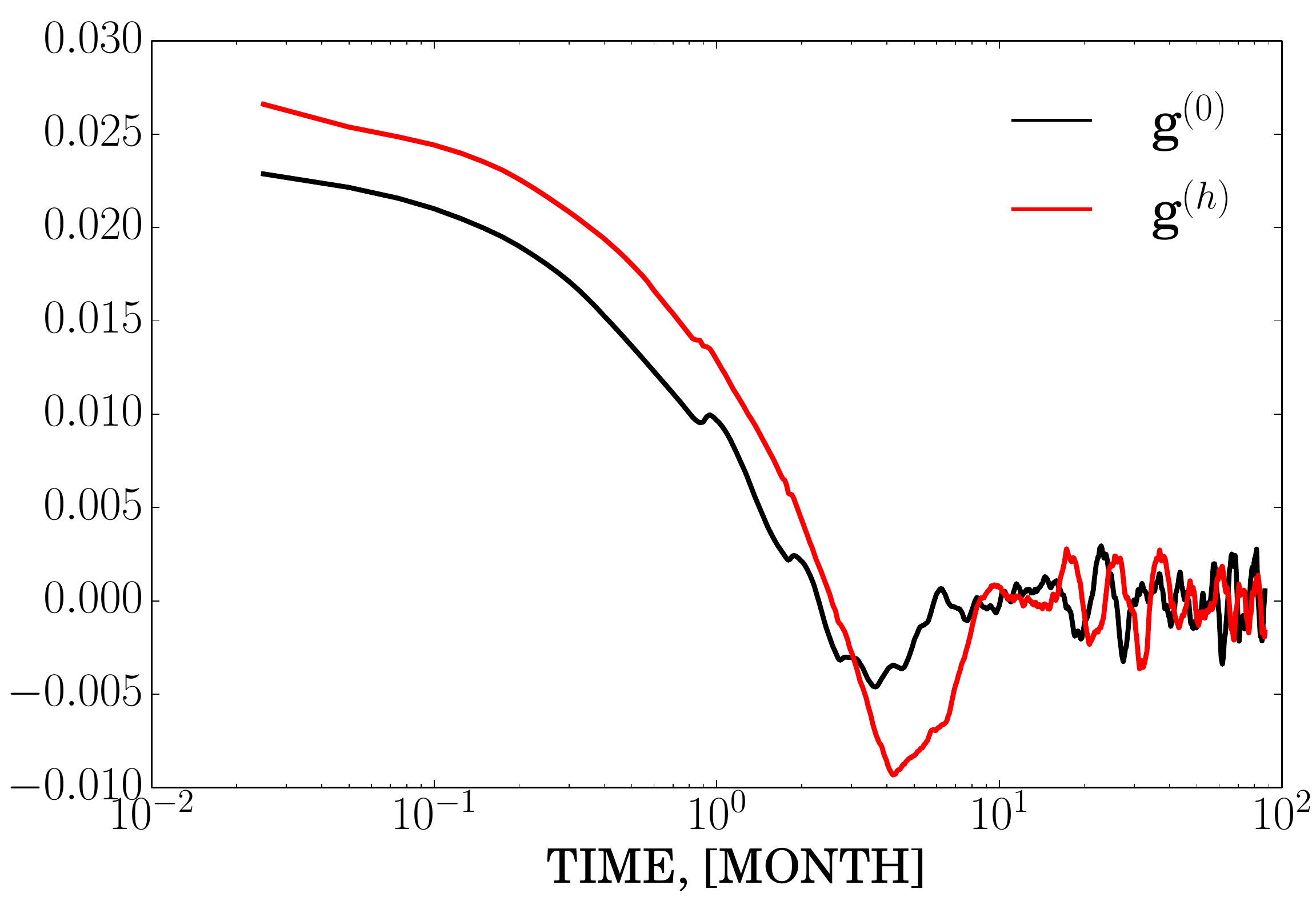} 
\caption{The auto-correlation function of the magnetic dipole fluctuations  (mean values are subtracted)}
\label{corr} 
\end{figure}

\label{resul}

\subsection{Axial dipole data}

\label{axialdip}

We calculate the rms value of the noise $\delta d$, the amplitude of
the magnetic dipole cycle variations $\bar{d}$ and its ratio $\delta d/\bar{d}$,
both for the observations ($d$ is estimated for cycle 22) and
the model to get the following results. The observations give $\delta d=0.24$,
$\bar{d}=2.2$,  $\delta d/\bar{d}=0.11$. For the amplitude of the
observed equatorial dipole we get similar results (see
Fig.~\ref{orig}b and Fig.~\ref{datam}).
 Here magnetic field is
normalized to the solar radius $R$ and magnetic moment is measured
in units of Gauss $R^{3}$. For illustration, we compare with
{ one of  our models}
which  has a half period (corresponding to the 11 yr sunspot cycle)
of 0.176 in our dimensionless units, $dt_{{\rm inj}}=0.0015=dt_{{\rm dec}}$,
and 5 spots are simultaneously present (although by the time a spot
is removed from the simulation, it has been present for 5 decay times,
and so its field is essentially reduced to zero). The model gives $\delta d=0.00085$,
$\bar{d}=0.015$, $\delta d/\bar{d}=0.058$. Here magnetic field is
measured in units of the equipartition magnetic field. For an order of magnitude estimate, 
we assume that the magnetic field unit in the model corresponds to $10^3$ G 
(typical field strength of sunspots) and then normalize to the solar radius. 
This gives $\delta = 15$ Gauss $R^3$ and $\delta d = 0.85$ Gauss $R^3$. Taking 
into account the very crude 
nature of the model, this seems to be in reasonable agreement with observations.
{ In physical units, the injection interval $dt_{\rm inj}$ is about 0.094 yr
(ca 1.1 months), which we chose rather arbitrarily to be equal to
the decay time $t_{\rm dec}$.}

We conclude that the model produces fluctuations
of the magnetic dipole that are comparable to those found in the observations.
The general nature of the results was not very sensitive
to the choice of parameters{ , provided that the interval between injections
is not too long or very short, or that the injected field strength is not much 
larger or smaller than the maximum dipole field strength produced by 
the dynamo. 
The results are quite insensitive to $t_{\rm dec}$, provided that it is not
very long, or to the number of spots present provided that $dt_{\rm in j}$
and $dt_{\rm dec}$ are not very different in magnitude}.

We can estimate the relative duration of the reversal epoch $t^{*}$
as follows. If the cyclic behaviour is sinusoidal,
we can estimate the time during which $\bar{d}\sin t$ is lower than
$\delta d$. A simple calculation gives $\delta d/\bar{d}=0.06$,
and for an 11-year cycle $t^{*}\approx4$ month.

The autocorrelation function of the noise is presented in Fig.~\ref{corr}.
The correlation time of the fluctuations can be estimated from the
plot as $\tau_{{\rm cor}}\approx 0.18$ yr$=65 {\rm d} \approx 2$ months, i.e. about half 
the duration of the reversal epoch. Note, that parameters of
  fluctuations of the axial dipole and the amplitude of the equatorial
dipole are rather similar.


We estimate $b/(B\sqrt{N})B_P/B\approx\delta d/\bar{d}$ {
(here $B$ is the mean magnetic field field and $B_P$ its poloidal component)}
following Eq.~(1) of Moss et al. (2013),
where $B$  is the large-scale magnetic field strength in the near-surface
layers within  the
Sun, $b$  is the corresponding strength of magnetic fluctuations and
$N$  is number of convective cells in the domain of dynamo action.
Following Moss et al. (2013) we take as a crude estimate
$N=10^{3}$  { and $B_P/B \approx 0.1$} and arrive at an estimate $b/B=2$,
which is more or less in agreement with the estimate obtained by Sokoloff
\& Khlystova  (2010) from statistics of the sunspot groups that violate
the Hale polarity law.

\subsection{The direction of the magnetic dipole at the epoch of reversal}

\begin{figure}
a)\includegraphics[width=0.85\columnwidth]{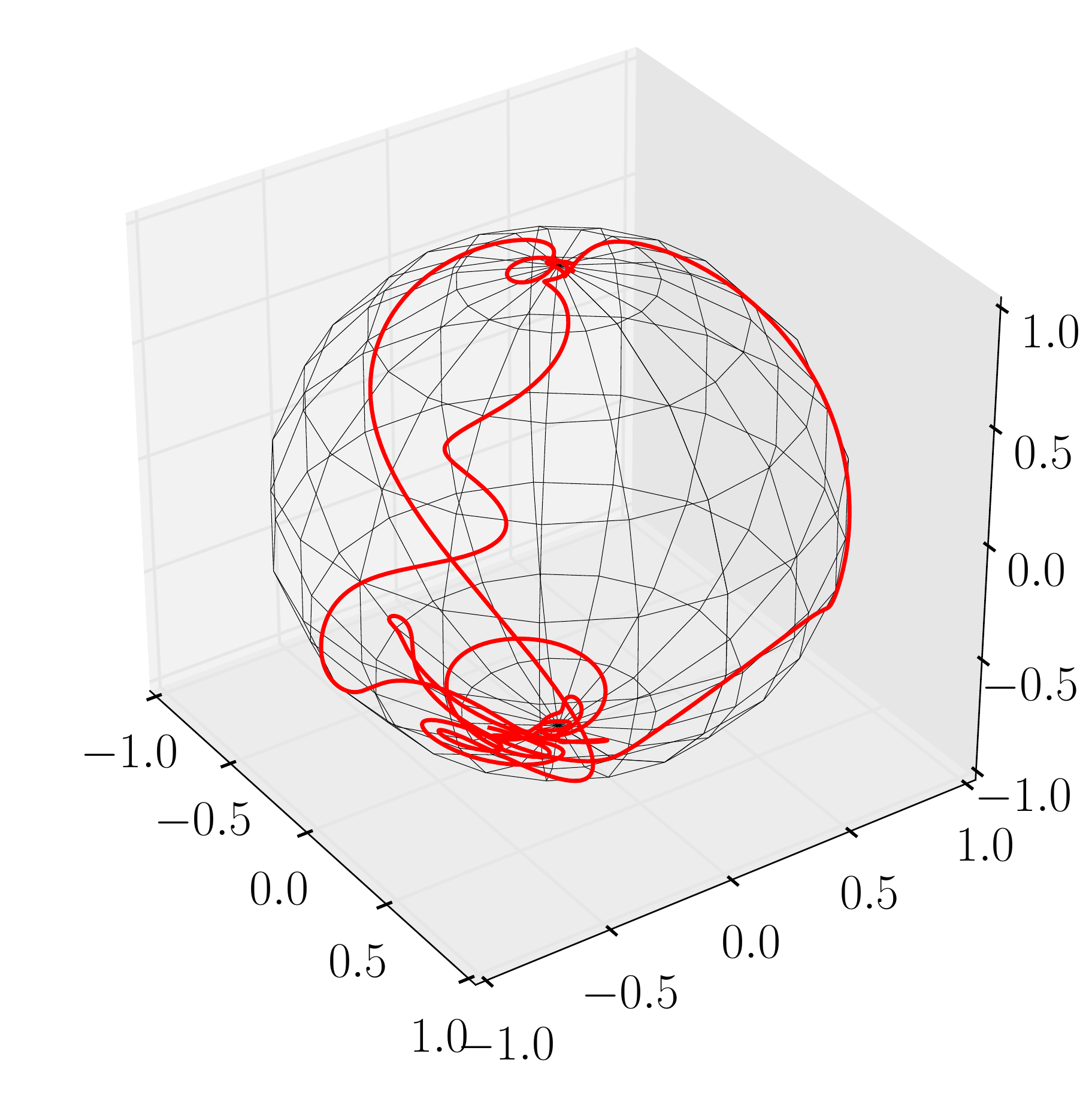}\\
b)\includegraphics[width=0.85\columnwidth]{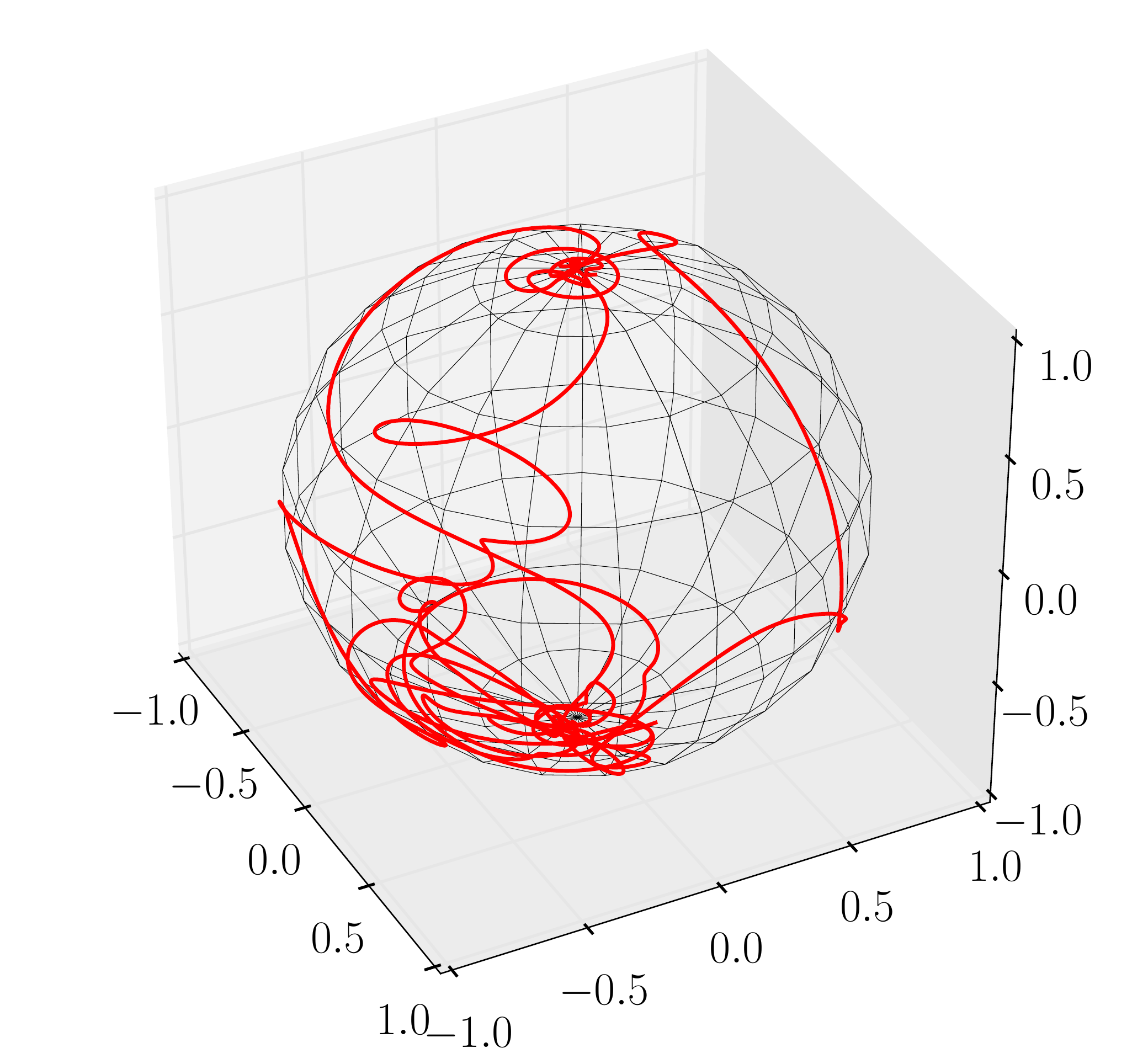}\\
c)\includegraphics[width=0.85\columnwidth]{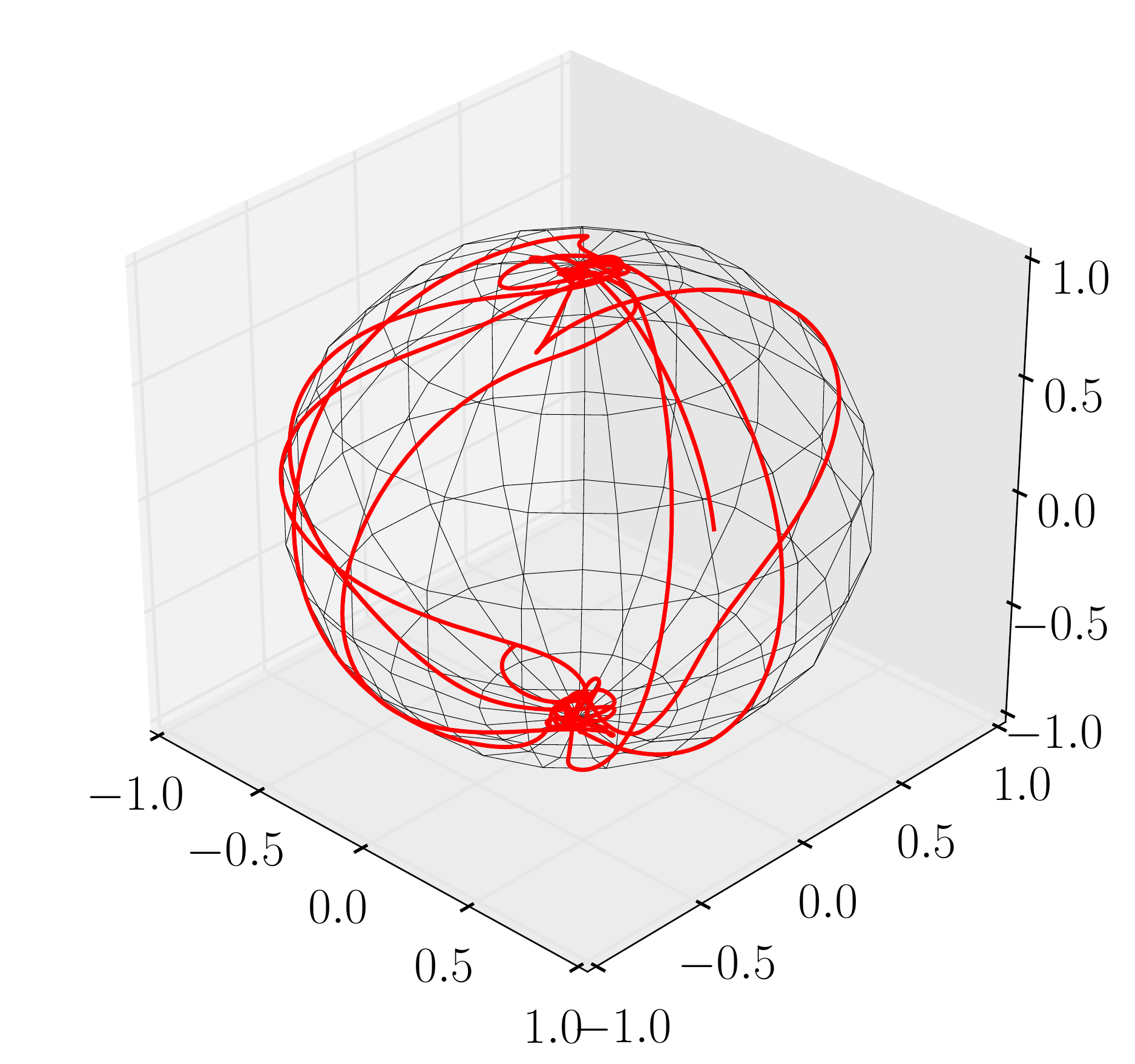}
\caption{ The track of the { pole the of the magnetic dipole} on a sphere of unit radius.
a) WSO data using a Gaussian filter with a one
year window for the model described in Fig.~\ref{mod}; b) the same for  the half year window; c) the same for
the model -- see Fig.~\ref{mod} -- which, in contrast to the data, covers an interval with 6
reversals of the global dipole.}
\label{inclin} 
\end{figure}

\begin{figure}
a)\includegraphics[width=0.85\columnwidth]{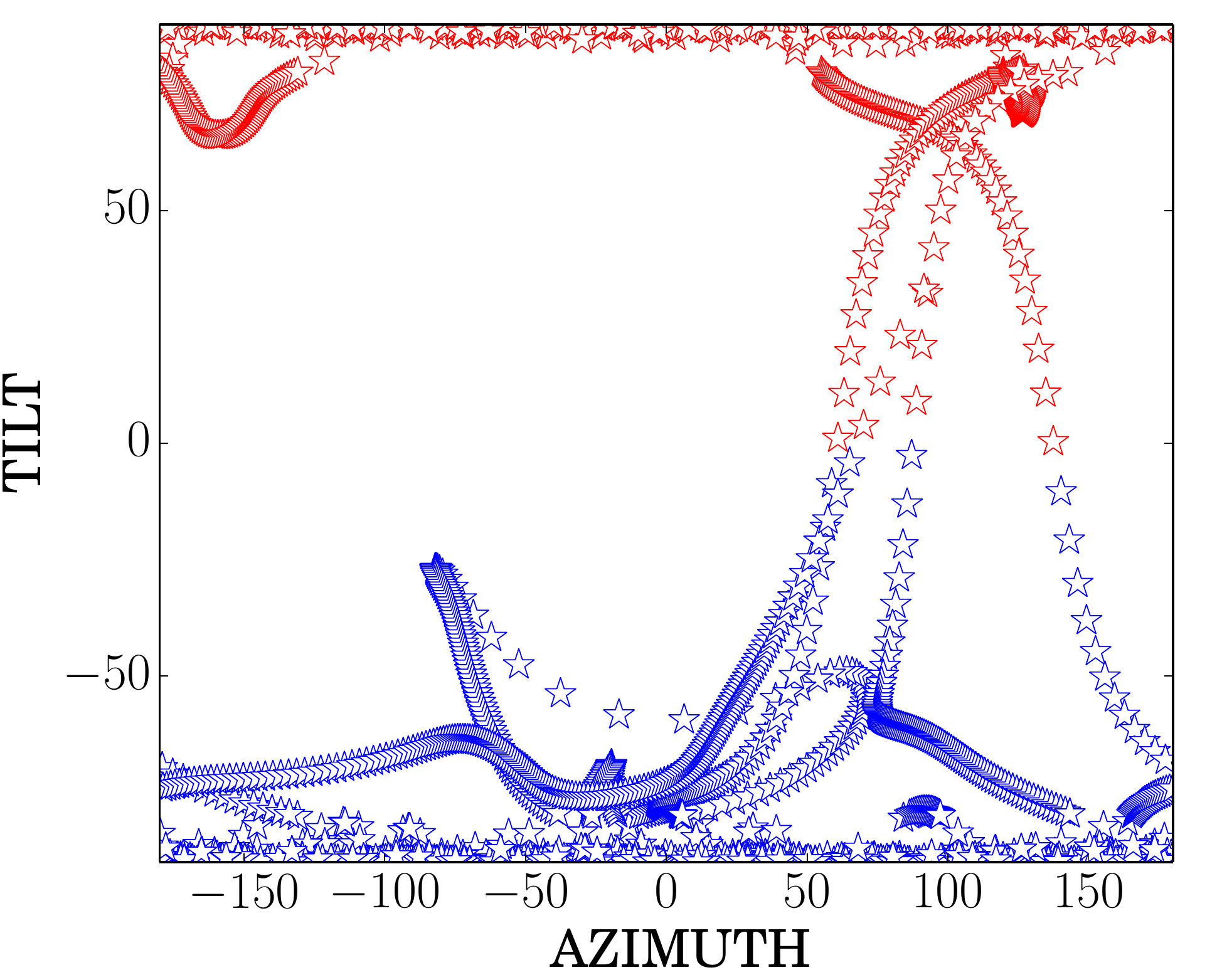}\\
b)\includegraphics[width=0.85\columnwidth]{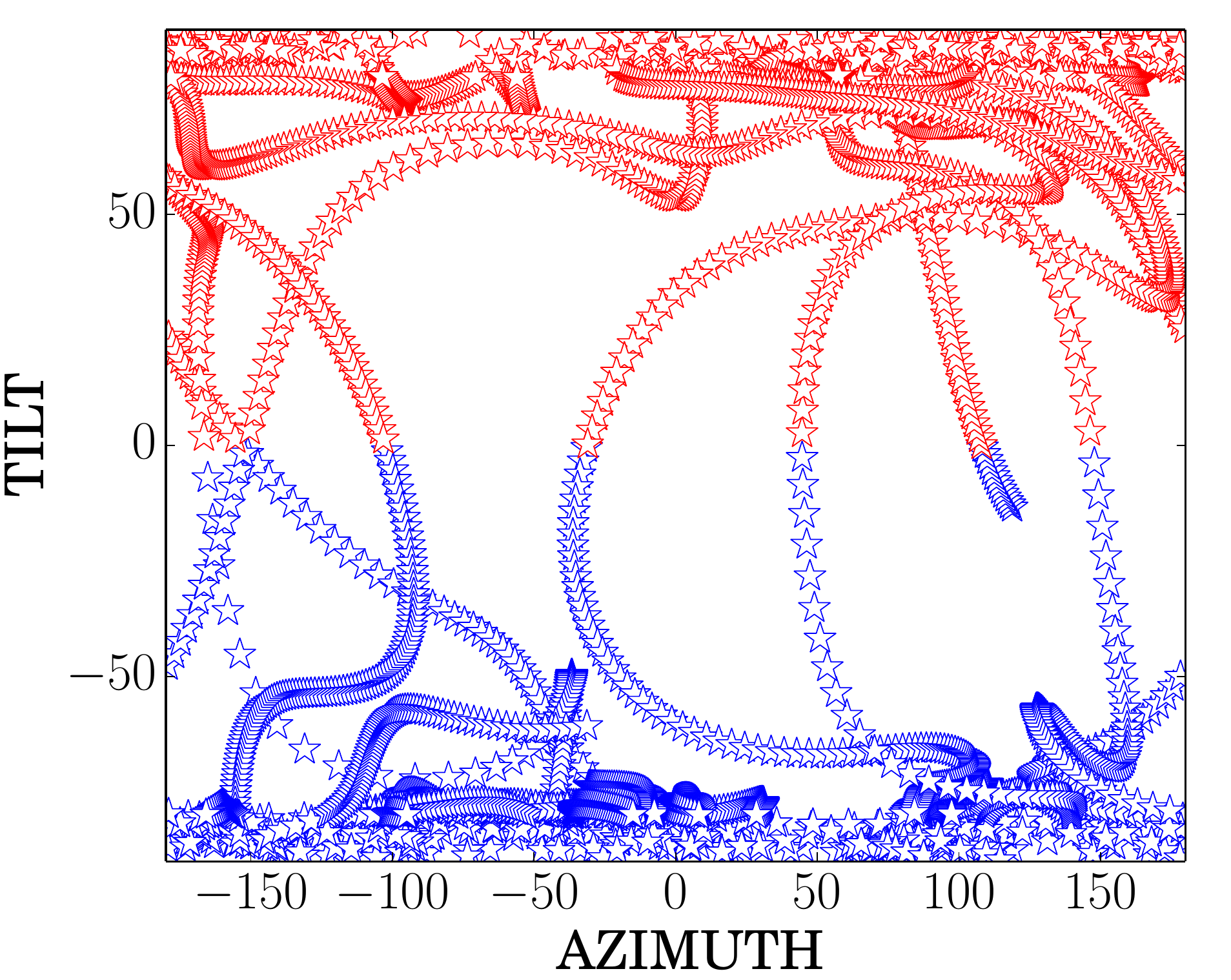}
\caption{The track of the { pole of the magnetic dipole} in the $\theta-\lambda$ plane:
a) WSO data, b) the same for
the model (Fig.~\ref{mod}) which, { in contrast to the 4 reversals in the
WSO data}, covers an interval with 6 reversals of the global dipole.}
\label{thetalamb} 
\end{figure}

We can follow the evolution of the total magnetic dipole during the
epoch of reversal by plotting the { position of the}
end of the dipole vector as a point on a sphere
of unit radius (Fig.~\ref{inclin}a). We conclude that the reversal
track for the model (Fig.~\ref{inclin}c) is quite similar to that deduced from observations.
Of course, a shorter filtering time gives a more complicated trajectory
for the dipole. { We give an alternative presentation of the dipole
track in Fig.~\ref{thetalamb}.}

We find that the trajectories of the { pole of the magnetic dipole} depend on the 
method of smoothing applied to 
the WSO data as well as for the model data.
These plots are quite similar to those shown in 
Fig.~\ref{inclin}b 
and we do not present them here. Using the model data, we demonstrated 
that the longitude  of reversal tracks is distributed on the
sphere quite randomly. 
Unfortunately, there are no corresponding long-term observational data to confirm
this point observationally.

\subsection{Cyclic evolution of the amplitude of magnetic fluctuations}

According to the interpretation of the observational data suggested, 
the equatorial magnetic dipole
is a large-scale manifestation of magnetic fluctuations.
A notable feature of the plots describing the
 equatorial dipole (Figs.~\ref{orig},\ref{transf}) is that the amplitude of 
the noise
is modulated with the solar cycle period. We did not include such
a feature in our model, so quite naturally the plot for the equatorial dipole 
for the model data (Fig.~\ref{mod})  does not show such an effect. 
If we include modulation of the fluctuations with the amplitude of
 the toroidal field variations, 
the desired modulation of the equatorial dipole appears.

We conclude that the rms of the solar magnetic fluctuations has an
11-year periodicity.  We find that this modulation
survives if we introduce artificially a long-term evolution 
of the cycle amplitude, as has occurred during recent solar cycles (we omit the plot for the sake of brevity).
We note that cyclic behaviour of the equatorial dipole was mentioned
previously by Hoeksema (1995) and Obridko \& Livshits (2006).

\section{Discussion and conclusions}
\label{disc}

Our analysis of the observational data,
and comparison with a simple model of the solar dynamo, confirm that non-zero
nonaxisymmetric values of the total magnetic dipole at the epoch
of solar magnetic reversal are consistent with a large-scale contribution
from { fluctuations of the solar surface magnetic field}, 
as discussed by Moss, Kitchatinov \& Sokoloff (2013).
{ We envisage that these fluctuations, {\bf connected loosely} to active regions, 
are driven by an instability of the underlying layers.}

The analysis gives an estimate for the ratio of the rms 
value of the fluctuation to the
amplitude of cyclic variations of the mean solar field of $b/B \approx 2$. 
This estimate
is independent of the estimate of Sokoloff \& Khlystova (2010), obtained from 
the statistics of  sunspot groups that do not follow the Hale polarity 
law.
These estimates differ 
by a factor of about 2, which appears quite  
satisfactory given that the amplitude of solar cycle varies substantially 
from cycle to cycle,  and that the estimates come from analyses
of quite different data.

The random nature of the orientation of the  solar magnetic dipole during the epoch of 
reversal is clearly visible from the trajectories
of the { pole of the magnetic dipole} on the unit sphere  (Fig.~\ref{inclin}c). On
the other hand,
the epoch of reversal occupies a relatively small part of the solar cycle 
(about 4 months only). The duration of the 
reversal epoch is estimated as the time during which the fluctuating part 
of the dipole is larger than the mean-field
contribution. 
In comparison, the memory time for the  dipole fluctuations is estimated 
as about 2 months, i.e.
about half the duration of the reversal epoch. This is why the dipole trajectory 
is far from just being noise,
and some traces of regular behaviour are visible in the trajectory  during a 
given reversal. This might be instructive 
for understanding of the phenomenon of active longitudes 
(e.g. Usoskin et al. 2007  and references therein). 

Our analysis gives further important information relevant to studies of the
solar dynamo. We see that magnetic fluctuations are cyclically modulated.
This means that they (at least mainly) originate in the periodic  global
dynamo action, rather than from small-scale dynamo action, as 
was suggested by Goode et al. (2012) and Abramenko (2013). Possibly
this is because active regions are generated more readily in the parts
of the cycle when toroidal fields are stronger -- recall that the
axisymmetric poloidal field is out of phase with the dynamo generated
toroidal field. In other words, we confirm here the interpretation of Stenflo (2013)
that a local dynamo does not give a significant contribution to the observed 
{small-scale} flux.
Following this interpretation however, based on other observational data, 
we infer that, as far as we
can confirm from the observations available, solar magnetic activity 
is a result of a single physical process
which simultaneously generates the solar activity cycle as well as
the  magnetic fluctuations. We do not exclude in principle the
possibility that an additional dynamo excitation mechanism for 
small-scale magnetic field is active somewhere in
the solar interior.  However we are unable to isolate its manifestations from  the
general cyclic behaviour associated with the
main driver of solar magnetic activity.

One more important point is that the small-scale magnetic fluctuations 
considered here result in weak 
nonaxisymmetric magnetic field components, as demonstrated by the
equatorial dipole data (Figs.~\ref{orig}, \ref{transf}). An 
additional illustration of this fact is given in Fig.~\ref{datam} 
which plots the large-scale total dipole component of the
solar activity, including both the axisymmetric and nonaxisymmetric
parts.
We see that the nonaxisymmetric 
components are stronger during maxima of sunspot activity, 
i.e. they are associated with the toroidal magnetic 
field. The lifetime of bursts of nonaxisymmetric components is of order 1 year. 
Any such large-scale components 
could be described by standard mean-field dynamo equations. 
Moss et al. (2013) estimated the lifetime 
of such nonaxisymmetric bursts as about 1 year, in reasonable
agreement with observations. 

Our interpretation is that the nonaxisymmetric magnetic field components appear 
as a result of decay of the large-scale toroidal magnetic field. 
{ The typical correlation time of the nonaxisymmetric modes is
estimated to be around 1 year. This agrees with results of the recent analysis of rotation
of solar active regions made by \cite{pelt10}. They, following \cite{KR80}, suggested  
the existence of non-oscillatory  nonaxisymmetric modes which rotate rigidly
with an angular velocity that is different from the overall rotation
period. These modes are coupled to the global toroidal field and
affected by differential rotation. This is consistent with the results
presented in Fig.~\ref{spek}. Note that wavelet analysis by
\cite{MPl00} reveals that nonaxisymmetric modes with different
periods of rotations are excited during different epochs of the
solar cycle, e.g. the mode with P=26.9 days is present during the
decaying phase of activity, while modes with periods around 28
days are present during the rising phase of the solar cycle (cf. Fig.~\ref{transf}b).}
Perhaps an equally  valid interpretation is to link the rotation rates
to the rotation rates of the remnant flux patterns.
A more complicated 
approach might include the possibility of azimuthal dynamo wave excitation 
in spherical shell convection  (Cole et 
al. 2013). Note, the rotational periods of the emerging nonaxisymmetric fields
can be
related to the ``new magnetic flux'' which might originate in a subsurface
shear layer -- \citet{Beal1999}. This is compatible with a distributed
dynamo operating in the bulk of the convection zone and being shaped by the
subsurface shear layer (\citealt{{bran05},{pk11}}). 
  
Obviously, our interpretation does not exclude the possibility
that occasionally the
equatorial magnetic dipole might vanish simultaneously with the
strength of the axial dipole
(which is part of the mean field) passing through zero during the 
course of an reversal. In such a case,  at some instant during the
reversal the total dipole will vanish. Remarkably (cf. Fig. \ref{datam}), 
this seems to be happening just now (Obridko 2013).
 Of course, there are details of the data that we have not analyzed,
which could affect our conclusions. These include, for example, the
study of effects of the higher
order components of the decomposition Eq(1) on reversals of polar
field. However, such a study goes beyond the simple dynamo model
discussed in this paper. 

{ In summary, we have presented a simple heuristic model -- basically
a cartoon -- to illustrate a 
mechanism to reproduce the behaviour of the solar dipole during reversals.
We have necessarily made a number of more-or-less arbitrary -- but reasonable --
assumptions. Our principle conclusion is that a model of this sort
can quite satisfactorily represent the observed phenomena during a reversal.
The mechanism seems quite robust, and quantitatively supports the idea
proposed by Moss, Kitchatinov \& Sokoloff (2013). Thus we feel confident that it captures
the essence of the solar behaviour without performing excessive fine tuning. 
Moreover, we find that the detailed observational data
concerning solar dipole reversals have, perhaps
 rather unexpectedly, the potential to reveal much about 
dynamo action in the solar interior. Clearly, there is considerable scope for 
a more detailed analysis of a more sophisticated model.}

\begin{acknowledgements}
V.P. and D.S. are grateful to RFBR for financial support under grant 12-02-00170-a. Useful discussions with V.N.Obridko are acknowledged.
\end{acknowledgements}


\end{document}